\documentclass[11pt,onecolumn]{article}
\usepackage[utf8]{inputenc}
\usepackage{amsmath,amsfonts,amssymb,graphicx,authblk,placeins,cite}
\usepackage{comment}
\usepackage{color,soul}
\usepackage{hyphenat}
\usepackage[top=50pt,bottom=50pt,left=70pt,right=70pt]{geometry}
\usepackage{sectsty}
\usepackage[font=sf]{caption}

\begin{document}
\allsectionsfont{\sffamily}
\pagenumbering{gobble}



\title{\sffamily \huge Predictive coding and stochastic resonance as fundamental principles of auditory perception}


\author[1,2]{\sffamily Achim Schilling}
\author[3]{\sffamily William Sedley}
\author[2,4]{\sffamily Richard Gerum}
\author[1,5]{\sffamily Claus Metzner}
\author[1]{\sffamily Konstantin Tziridis}
\author[6]{\sffamily Andreas Maier}
\author[1]{\sffamily Holger Schulze}
\author[7]{\sffamily Fan-Gang Zeng}
\author[8]{\sffamily Karl J. Friston}
\author[1,2,6]{\sffamily Patrick Krauss}


\affil[1]{\small{Neuroscience Lab, University Hospital Erlangen, Germany}}

\affil[2]{\small{Cognitive Computational Neuroscience Group, University Erlangen-N\"urnberg, Germany}}

\affil[3]{\small{Translational and Clinical Reseearch Institute, Newcastle University Medical School, Newcastle upon Tyne NE2 4HH, UK}}

\affil[4]{\small{Department of Physics and Center for Vision Research, York University, Toronto, Ontario, Canada}}

\affil[5]{\small{Biophysics Lab, University Erlangen-N\"urnberg, Germany}}

\affil[6]{\small{Pattern Recognition Lab, University Erlangen-N\"urnberg, Germany}}

\affil[7]{\small{Center for Hearing Research, Departments of Anatomy and Neurobiology, Biomedical Engineering, Cognitive Sciences, Otolaryngology–Head and Neck Surgery, University of California Irvine, Irvine, California 92697, USA}}

\affil[8]{\small{Wellcome Centre for Human Neuroimaging, Institute of Neurology, University College London, London WC1N 3AR, UK}}

\maketitle


\begin{abstract}{\sffamily \noindent
How is information processed in the brain during perception? Mechanistic insight is achieved only when experiments are employed to test formal or computational models. In analogy to lesion studies, phantom perception may serve as a vehicle to understand the fundamental processing principles underlying auditory perception. With a special focus on tinnitus -- as the prime example of auditory phantom perception -- we review recent work at the intersection of artificial intelligence, psychology, and neuroscience. 
In particular, we discuss why everyone with tinnitus suffers from hearing loss, but not everyone with hearing loss suffers from tinnitus. We argue that the increase of sensory precision due to Bayesian inference could be caused by intrinsic neural noise and lead to a prediction error in the cerebral cortex. Hence, two fundamental processing principles - being ubiquitous in the brain - provide the most explanatory power for the emergence of tinnitus: predictive coding as a top-down, and stochastic resonance as a complementary bottom-up mechanism. 
We conclude that both principles play a crucial role in healthy auditory perception.

}
\end{abstract}

\newpage
\section*{Introduction}
The ultimate goal of neuroscience is to gain a mechanistic understanding how information is processed in the brain. 
Since the early beginnings of the scientific study of the brain, lesions or more broadly anatomical damages and their physiological effects have provided pivotal insights into brain function.  
Analogously, phantom perception may serve as a vehicle to understand the fundamental processing principles underlying perception. 
The prime example of an auditory phantom perception is tinnitus, which is caused by anatomical damages along the auditory pathway.
Here we provide a mechanistic explanation of how tinnitus emerges in the brain: namely, how the neural and mental processes underlying perception, cognition, and behavior contribute to—and are affected by—the development of tinnitus. These insights may not only point to strategies, how tinnitus may be reversed or at least mitigated, but also how auditory perception is implemented in the brain in general. 
\\
While there is broad agreement in the scientific community on these goals, there is far less agreement on the way to achieve them. There is still a popular belief among neuroscientific and psychological tinnitus researchers that we are largely data driven. In other words, generating large, multi-modal, and complex data sets—analyzed with advanced machine learning algorithms—will lead to fundamental insights into how tinnitus emerges. Indeed, in the last decades we have assembled a broad data base, which has inspired models that make quantitative predictions. These predictions scaffold new experimental paradigms that aim to unravel the mechanisms of tinnitus perception. In the following, we summarize some of the main findings in tinnitus research, over the last decades, and then turn to strategic questions about how to leverage these advances, from the perspective of formal modelling.

Some universal correlations between hearing loss, tinnitus, and neural hyperactivity in the auditory system have been found in both animal and human studies. These reproducible findings can be considered as the common denominator of tinnitus research, and could offer the minimal starting point for theoretical considerations. Tinnitus is a phenomenon arising somewhere along the auditory pathway, but not in the inner ear \cite{schaette2012computational}. Thus, it can be shown that the spontaneous activity of neurons along the auditory pathway is increased after hearing loss \cite{kaltenbach2000hyperactivity, kaltenbach2002cisplatin, kaltenbach2006dorsal}, whereas the damaged cochlea transmits less information to the higher auditory nuclei \cite{moore1996perceptual, tziridis2021tinnitus}. However, it has been argued that not all alterations in neural activity in animal models, which were caused by an acoustic trauma, are necessarily related to tinnitus \cite{schaette2012computational, eggermont2013hearing}. Although there exist some behavioral tests to check for the putative presence of a tinnitus percept based on conditioning \cite{jastreboff1988phantom, lobarinas2004novel} or startle responses \cite{turner2006gap, schilling2017new, gerum2019open}, the reliability of these paradigms remains controversial \cite{eggermont2013hearing,eggermont2015tinnitus}. Thus, studies on human subjects complement these findings. In several recent studies, it was shown that the tinnitus frequency lies within the range of the hearing loss \cite{yakunina2021does, keppler2017relationship, dalligna2014there, schecklmann2012relationship}. Furthermore, Dalligna and coworkers report that the tinnitus is directly centered at the frequency of the largest hearing loss \cite{dalligna2014there} and Keppler and coworkers report that there is no correlation between tinnitus pitch and the edges of the impaired frequency ranges \cite{keppler2017relationship}. Nevertheless, there are also some studies reporting a correlation between the edges of the hearing loss and the tinnitus pitch \cite{konig2006course, moore2010relationship, pan2009relationship}.

Indeed, the above neural correlates of tinnitus and hearing loss are just a small distillation of all studies that aspire to unravel mechanisms that underwrite tinnitus, but these findings are robust and constitute the basis of most theoretical and computational models of tinnitus. In the 1990s the first computational models of tinnitus emerged. These models considered decreased lateral inhibition—due to deficient auditory input—as the main cause of tinnitus. Gerken \cite{gerken1996central} created a feed-forward brainstem model and suggested the inferior colliculus to be the crucial structure for tinnitus development. Kral and Majernik \cite{kral1996lateral}, as well as Langner and Wallh\"auser-Franke \cite{langner1999computer} pursued computational models, based on decreased lateral inhibition. Bruce and coworkers developed these models further and implemented lateral inhibition in a spiking recurrent neural network \cite{bruce2003lateral}. In a subsequent step, the principles were implemented in a model of the auditory cortex based on spiking neurons \cite{dominguez2006spiking}.

Besides lateral inhibition, homeostatic plasticity \cite{turrigiano1999homeostatic} and central gain changes are hypothesized to be the cause for tinnitus emergence and manifestation. Thus, Parra and Pearlmutter developed an "abstract" model, where they simply defined several frequency channels with a certain input \cite{parra2007illusory}. The output was scaled with the average, which means that a decreased input leads to a higher scaling factor. However, they did not consider a plausible neural implementation of this mathematical model. Schaette and Kempter further developed several computational models, investigating the effects of central gain increase on tinnitus emergence \cite{schaette2006development, schaette2008development, schaette2009predicting}. Finally, Chrostowski and coworkers developed a cortex model to investigate central gain changes in the cortex  (\cite{chrostowski2011can}, for detailed review on computational tinnitus models see e.g. \cite{schaette2012computational}).

In 2013, Zeng introduced a model that argues that tinnitus is not caused by increased central gain, but by increased central noise, which means an additive neural noise, that is intrinsically generated \cite{zeng2013active}. The idea of an additional intrinsic or extrinsic noise as an explanation for tinnitus has gained some popularity in recent years (see e.g., Koops and Eggermont \cite{koops2021thalamus}). Zeng raised the question why the brain should increase central noise levels. This question was addressed in 2016 by Krauss and coworkers \cite{krauss2016stochastic}, who showed that internally generated neural noise could partially restore hearing ability after hearing loss through the effect of stochastic resonance \cite{krauss2016stochastic, krauss2017adaptive, krauss2018cross, schilling2021stochastic}. This hypothesis is supported by the finding that on average hearing thresholds are better in patients suffering from hearing loss with tinnitus compared to a control group of patients suffering from hearing loss but without tinnitus \cite{gollnast2017analysis}. Along the same line, the stochastic resonance and central noise model may explain the Zwicker tone illusion \cite{zwicker1964negative}, i.e. the perception of a phantom sound which occurs after stimulation with notched noise, and why auditory sensitivity for frequencies adjacent to the Zwicker tone are improved beyond the absolute threshold of hearing during Zwicker tone perception \cite{wiegrebe1996auditory}. Furthermore, recently, a crucial prediction of the stochastic resonance model of tinnitus development was confirmed experimentally by using brainstem audiometry \cite{schilling2019objective} and assessing behavioral signs of tinnitus \cite{gerum2019open} in an animal model: simulated transient hearing loss improves auditory sensitivity and leads, as a side effect, to the perception of tinnitus \cite{krauss2021simulated}. Both, the model from Zeng and the model from Krauss et al., are not based on a particular or specified neural network architecture. However, in 2020, Schilling and coworkers developed such a model based on a hybrid neural network of spiking neurons and a deep neural network, proving that intrinsically generated noise could indeed significantly increase speech perception via stochastic resonance \cite{schilling2020intrinsic}. Recently, a similar hybrid neural network model has led to further insights into the mechanisms of impaired speech recognition caused by hearing loss \cite{haro2020deep}. 

In parallel to the intrinsic neural noise models from Zeng and Krauss et al., Sedley and coworkers developed a conceptual model, which describes tinnitus as arising from a prediction error of the brain \cite{sedley2016integrative, sedley2019exposing}. This model is based on the idea that the brain is a Bayesian prediction machine, trying to minimize prediction errors or free energy \cite{friston2010free, friston2018does}, and therefore that modifying auditory predictions to accord with noise in the auditory pathway reduces prediction errors, but at the expense of causing an erroneous perception (i.e. tinnitus). More recently, a computational instantiation of this model, using a hierarchical Gaussian filter, has been able to explain phenomenology in individual tinnitus subjects, and their predict residual inhibition characteristics \cite{hu2021bayesian}. Furthermore, Dotan and Shriki introduced a recurrent neural network model where tinnitus-like “hallucinations” are elicited by sensory deprivation as a result of entropy maximization \cite{dotan2021tinnitus}. Finally, Gault and co-workers fitted audiometric data with assessed tinnitus pitches in order to create a linear regression model of tinnitus pitch and loudness \cite{gault2020perceptual}. Besides these computational models that rest upon a mathematical formulation, there exist several phenomenological models, such as the thalamo-cortical dysrhythmia model \cite{llinas1999thalamocortical, de2015thalamocortical}, the thalamic low-threshold calcium spike model \cite{jeanmonod1996low}, the frontostriatal gating hypothesis \cite{rauschecker2015frontostriatal, knipper2020neural}, and the overlapping sub-network theory \cite{de2011phantom, vanneste2012auditory}. Finally, there exist another Bayesian brain / predictive coding model of tinnitus, which is kind of the polar opposite to what Sedley and Friston have argued for. There, tinnitus is not believed to arise from spontaneous noise increase, which higher predictions go on to accept, but on the contrary that tinnitus arises from reduced input to the auditory cortex, leading it to ‘make up’ or ‘fill in’ an auditory percept from auditory memory \cite{de2014bayesian}. However, this assumption contradicts the findings that neural activity is increased after hearing loss \cite{kaltenbach2000hyperactivity,kaltenbach2002cisplatin,kaltenbach2006dorsal}.

As there exist various models of tinnitus development, criteria are needed to define which models are apt to understand tinnitus development. In their review paper Schaette and Kempter \cite{schaette2012computational} define three major criteria for the quality of a model: First, -- and in line with Popper's ideas \cite{popper1963science} -- a model should be falsifiable, which means there should be experimental paradigms, which could be used to test whether a model is right or wrong. Second, a model should make quantitative predictions, as opposed to purely qualitative, often vague, predictions (compare also \cite{lazebnik2002can}). Third, a model should be as simple as possible, i.e., contain the smallest number of parameters and assumptions as possible, a principle called Ockham’s razor \cite{lazar2010ockham}. Hence, when two models explain experimental data equally well, the simpler one is considered to be the better one.

With the huge progress of artificial intelligence (AI) during the last decade, which is mainly due to novel hardware resources, a new discipline has been founded, called \emph{Cognitive Computational Neuroscience} (CCN) as an integrative endeavor at the intersection of AI, cognitive science and neuroscience \cite{kriegeskorte2018cognitive, naselaris2018cognitive, kietzmann2019deep}.

In this paper, we first discuss the opportunities and limitations of this new research agenda. In particular, we present key thought experiments that highlight the major challenges on the road towards a CCN of tinnitus. In the light of these considerations, we subsequently review current models of tinnitus and assess their explanatory power. Finally, we present an integration of those models that we consider most promising, and point towards a unified theory of tinnitus development.

\section*{Three challenges ahead}

\subsection*{The challenge of developing a common formal language}

In 2002, Yuri Lazebnik compared the biologists’ endeavor -- of trying to understand the building blocks and processes of living cells -- with the problems that engineers typically deal with \cite{lazebnik2002can}. In his opinion paper \emph{Can a biologist fix a radio? -- Or, what I learned while studying apoptosis}, Lazebnik argued that many fields of biomedical research at some point reach \emph{``a stage at which models, that seemed so complete, fall apart, predictions that were considered so obvious are found to be wrong, and attempts to develop wonder drugs largely fail. This stage is characterized by a sense of frustration at the complexity of the process''} \cite{lazebnik2002can}.

Subsequently, Lazebnik discussed a number of intriguing analogies between the physical and life sciences. In particular, he identified formal language as the most important difference between the two. Lazebnik argues that biologists and engineers use very different languages for describing phenomena. On the one hand, biologists draw box-and-arrow diagrams, which are -- even if a certain diagram makes overall sense -- difficult to translate into quantitative assumptions, and hence limits its predictive or investigative value. 

Indeed, these thoughts fit to the criterion for a ‘good model’ as pointed out by Schaette and Kempter \cite{schaette2012computational}, ], i.e., that a model should make quantitative predictions. However, on the other hand a model should be as simple as possible and understandable, which means that it is important to find a compromise between too fine-grained and too coarse-grained descriptions of the system (see Marr's levels of analysis \cite{marr1979computational}, Fig. \ref{descriptionLevels_1} a). 

Lazebnik also remarks that scientific assumptions and conversations are often \emph{``vague''} and \emph{avoid clear, quantifiable predictions''} \cite{lazebnik2002can}. A freely adapted example drawn from Lazebnik's paper would be a statement like \emph{``an imbalance of excitatory and inhibitory neural activity after hearing-loss appears to cause an overall neural hyperactivity, which in turn seems to be correlated with the perception of tinnitus.''}

Descriptions of electrophysiological findings are and important starting point for hypothesis generation, but they are no more than a first step. Description needs to be complemented with explanation and prediction (compare also the four main goals of psychology as described e.g., in \cite{holt2019ebook}).

Furthermore, Lazebnik urges a more formal common language for biological sciences, in particular a language which has the precision and expressivity found in engineering, physics, or computer science. Any engineer trained in electronics for instance, is able to unambiguously understand a diagram describing a radio or any other electronic device. Thus, engineers can discuss a radio using terms that are common ground in the community. Furthermore, this commonality enables engineers to identify familiar functional architectures or motifs; even in a diagram of a completely novel device. Finally, due to the mathematical underpinnings of the language used in engineering, it is perfectly suited for quantitative analyses and computational modeling. For instance, a description of a certain radio includes all key parameters of each component like the capacity of a capacitor, but not irrelevant parameters -- that do not ‘matter’ -- like its color, shape, or size.

We emphasize that this does not mean that anatomical descriptions are useless in order to understand brain function. However, also in neurobiology there exist both kinds of detail: those that are crucial for understanding neural processing, and those that are not relevant variables.

Lazebnik concludes that \emph{``the absence of such language is the flaw of biological research that causes David’s paradox''}, i.e., the paradoxical phenomenon frequently observed in biology and neuroscience that \emph{``the more facts we learn the less we understand the process we study''} \cite{lazebnik2002can}. 

Some conclusions for tinnitus research can be drawn from Lazebniks’ thoughts on a more formal approach in biological sciences. The "central gain" and "homeostatic plasticity" theory on tinnitus emergence is a good example, how the communication on tinnitus research can be improved. For example, Roberts stated in 2018 that the increase of central gain is \emph{"increase of input output functions by forms of homeostatic plasticity''}, which means that homeostatic plasticity is necessarily connected to central gain adaptations \cite{roberts2018neural}. In contrast to that, Schaette and Kempter state that central gain changes can occur within seconds and thus are not necessarily caused by homeostatic plasticity \cite{schaette2012computational}. Only on longer time-scales both effects can be regarded as \emph{``functionally equivalent''} \cite{schaette2012computational}. Indeed, tinnitus research would profit from a unified terminology for the different constructs, in the best case, a mathematical formulation. 

\subsection*{The challenge of developing a unified mechanistic theory}

In 2014, Joshua Brown built on Lazebnik’s ideas and published the opinion article \emph{The tale of the neuroscientists and the computer: why mechanistic theory matters} \cite{brown2014tale}. In this thought experiment, a group of neuroscientists finds an alien computer and tries to figure out its function.

First, the M/EEG researcher tried to investigate the computer. She found that every time \emph{``when the hard disk was assessed, the disk controller showed higher voltages on average, and especially more power in the higher frequency bands''} \cite{brown2014tale}.

Subsequently, the cognitive neuroscientist, i.e., the fMRI researcher argued that M/EEG has insufficient spatial resolution to see what is going on inside the computer. He carried out a large number of experiments, the results of which can be summarized with the realization that during certain tasks, certain regions seem to be more activated and that none of these components could be understood properly in isolation. Thus, the researcher analyzed the interactions of these components, showing that there is a vast variety of different task-specific networks in the computer. 

Finally, the electrophysiologist noted, critically, that his colleagues may have found coarse-grained patterns of activity, but it is still unclear what the individual circuits are doing. He starts to implant micro-electrode arrays into the computer and probes individual circuit points by measuring voltage fluctuations. With careful observation, the electrophysiologist identifies units responding stochastically when certain inputs are presented, and that nearby units seem to process similar inputs. Furthermore, each unit seems to have characteristic tuning properties.

Brown’s tale ends with the conclusion that even though, they performed a multitude of different empirical investigations, yielding a broad range of interesting results, it is still highly questionable whether \emph{``the neuroscientists really understood how the computer works''} \cite{brown2014tale}.

This provocative thought experiment speaks to some ideas that are relevant for tinnitus research. In 2021, four leading scientists in tinnitus research discussed different tinnitus models at the Annual-Mid-Winter Meeting of the Association for Otolaryngology and diagnosed  a \emph{``lack of consistency of concepts about the neural correlate of tinnitus''} \cite{knipper2021too}. Thus, a clearly defined theoretical framework is needed, which helps empirical groups to develop experimental paradigms suited to verify or falsify alternative models. To achieve that, inter-disciplinary teams or at least an inter-disciplinary approach is needed \cite{silver2007neurotech}.


\subsection*{The challenge of developing appropriate analysis methods}

In 2017 Jonas and Kording implemented the thought experiment of Brown in an experimental study. In their study \emph{``Could a neuroscientist understand a microprocessor?''} \cite{jonas2017could} the authors address this question by emulating a classical microprocessor, the \emph{MOS 6502}, which was implemented as the central processing unit (CPU) in the Apple I, the Commodore 64, and the Atari Video Game System, in the 1970s and 1980s. In contrast to contemporary CPUs, like Intel’s \emph{i9-9900K}, that consist of more than 3 billion transistors, the \emph{MOS 6502} only consisted of 3,510 transistors, served as a ``model organism'' in the mentioned study, and performed three different ``behaviors'', i.e. three classical video games (Donkey Kong, Space Invaders and Pitfall).

The idea behind this approach is that the microprocessor, as an artificial information processing system, has three decisive advantages over natural nervous systems. Firstly, it is fully understood at all levels of description and complexity, from its gross architecture and the overall data flow, through logical gate primitives, to the dynamics of single transistors. Secondly, its internal state is fully accessible without any restrictions to temporal or spatial resolution. And thirdly, it offers the ability to perform invasive experiments on it. Using this framework, the authors applied a wide range of popular data analysis methods from neuroscience to investigate the structural and dynamical properties of the microprocessor. The methods used included -- but were not restricted to -- Granger causality for analyzing task specific functional connectivity, time-frequency analysis as a hallmark of M/EEG research, spike pattern statistics, dimensionality reduction, lesions, and tuning curve analysis.

The authors concluded that although each of the applied methods yielded results strikingly similar to what is known from neuroscientific or psychological studies, none of them could actually elucidate how the microprocessor works, or more broadly speaking, was appropriate to gain a mechanistic understanding of the investigated system.

There are potential criticisms of this study; for example, the brain is no computer and thus the drawn parallels are insufficient. Nevertheless, the idea to use a known model system to check for the validity of the evaluation procedures and common methods is a seminal principle. In 2009 Bennett and coworkers performed an even stranger experiment, when they used standard fMRI and statistics techniques to analyze the brain activity of a dead salmon, and indeed found a BOLD signal due to stimulation \cite{bennett2009neural}. At first glance this experiment seemed to be at least useless if not even funny, but it was a wake-up call and indeed changed the way fMRI data is evaluated. Nowadays there exist strict rules how to correct for multiple testing in fMRI research, to prevent pseudo-effects being a result of wrong statistical testing \cite{bennett2009neural, bennett2009principled}. In computational neuroscience and AI research, newly developed methods are always applied to standard data sets such as MNIST \cite{lecun1995learning, gerum2021integration} or artificially generated data sets with known properties (e.g. \cite{zenke2021remarkable, schilling2021quantifying, schilling2019objective}). The principle of using fully known -- even trivial systems -- to test the validity of tools, methods or even theories could be an important device in tinnitus research. Even in computational modeling, simply implementing a system in all details without a theory, which serves as a solid base, will not lead to a real understanding. Indeed, theory needs computational modeling, but the statement is also true the other way around \cite{gerstner2012theory}. Thus, computational models have to be checked in a sense that they should fulfill the minimum requirement -- that they can explain well understood and simple phenomena, as a back-check for further conclusions.

\section*{Towards a cognitive computational neuroscience of tinnitus}

\subsection*{What does it mean to understand a system?}

If popular analysis methods fail to deliver mechanistic understanding, what are the alternative approaches?  Most obviously, narrative hypotheses about the structure and function of the system under investigation will help. Instead of simply describing data features with correlations, coherence, Granger causality et cetera -- in the hope of learning something about the functioning of the system under investigation -- it would be much more effective to have a concrete hypothesis about the structure or function architecture of the system and then search for empirical evidence for that and alternative hypotheses.

Note that, this does not exclude explorative analysis of existing data, in order to generate new hypotheses. However, as we  pointed out in a previous publication \cite{schilling2021analysis}, in order to avoid statistical errors due to \emph{``HARKing''}\footnote{HARKING (\textbf{H}ypothesizing \textbf{a}fter \textbf{r}esults are \textbf{k}nown) is defined as generating scientific statements exclusively based on the analysis of huge data sets without previous hypotheses.} \cite{kerr1998harking, munafo2017manifesto} and to guarantee consistency of the results, it is necessary to apply e.g. re-sampling techniques such as sub-sampling \cite{schilling2019objective}, or alternatively, the well-established machine learning practice of cross-validation: i.e. splitting the data set into multiple parts before the beginning of the evaluation. One part is used for generating new hypotheses, and another part for subsequently statistically testing these hypotheses. Accumulation of such data-driven knowledge may finally even lead to a new theory.

Ideally, the verbally defined (narrative) hypotheses to be experimentally tested would be derived from such an underlying theory. As Kurt Lewin, the father of modern experimental psychology, pointed out: \emph{``There is nothing so practical as a good theory''} \cite{lewin1951field}. If we had theorized that the microprocessor from the thought experiment above performs arithmetic calculations, we could have derived the hypothesis, for example, that there must be something like 1-bit adders, and could have searched for them specifically. 

Conversely, Allan Newell, one of the fathers of artificial intelligence, stated that \emph{``You can‘t play 20 questions with nature and win''} \cite{newell1973you}. This suggests that testing one narrative hypothesis after another will never lead to a mechanistic understanding. So, this raises the fundamental question of what it actually means to ‘understand’ a system.  

Yuri Lazebnik argued that understanding of a system is achieved when one could fix a broken implementation: \emph{``Understanding of a particular region or part of a system would occur when one could describe so accurately the inputs, the transformation, and the outputs that one brain region could be replaced with an entirely synthetic component''} \cite{lazebnik2002can}. In engineering terms, this understanding can be simply described as $y=f(x)$, where $x$ is the input, $y$ is the output, and $f$ is the transformation. Fourier, Hilbert, rectifier, transformers etc. are all details; we can examine the input and output functions in frequency or time domains, in real or complex domains – these are all tools.

\begin{figure}[ht!]
	\centering
	\includegraphics{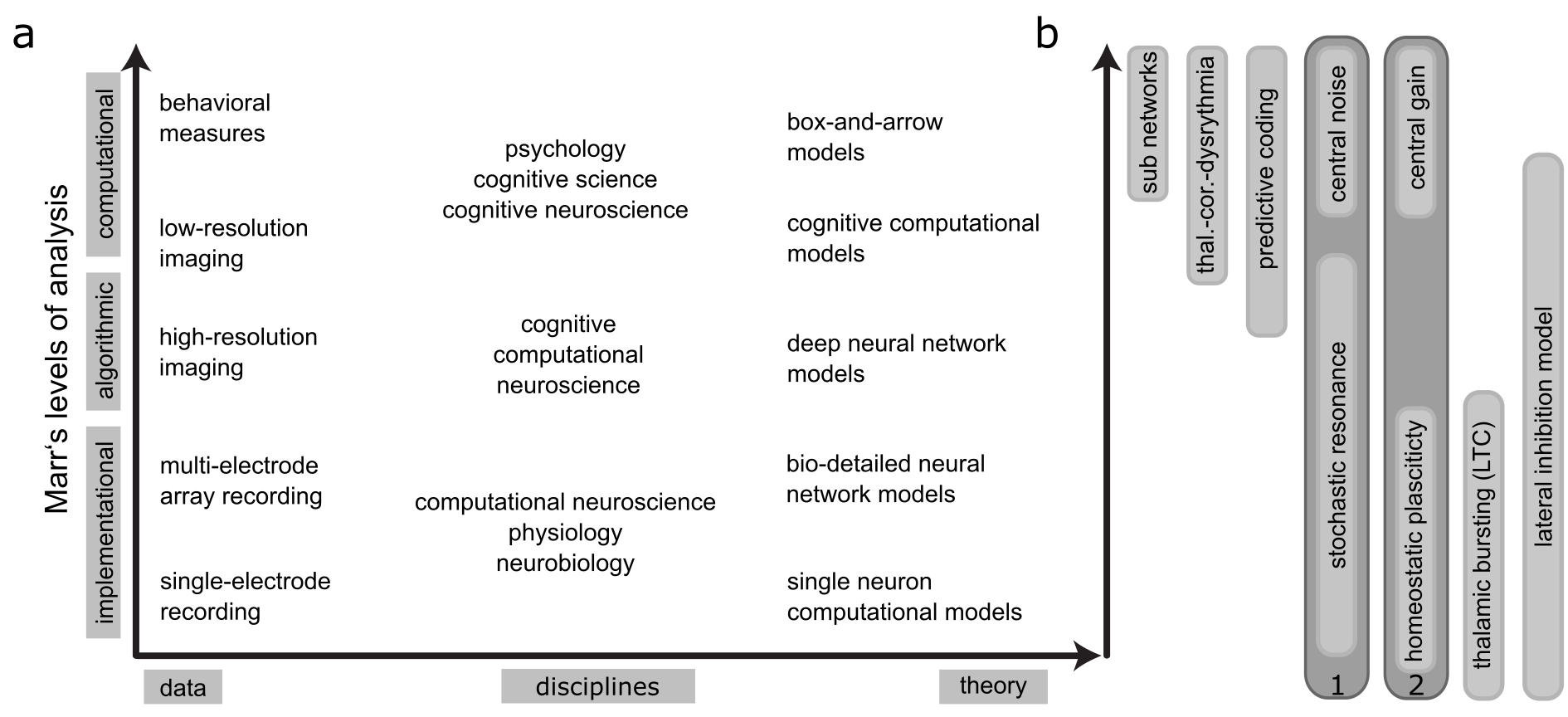}
	\caption{\textbf{Marr's Levels of Analysis:} a: The scheme illustrates how measurement methods (such as MEG, EEG, etc.), neuroscientific disciplines, as well as theoretical models can be structured in three different levels of analysis (according to Marr \cite{marr1979computational}). b: Tinnitus models in the light of the three levels of analysis; The gray bars illustrate how the different models cover the different levels of analysis (implementational, algorithmic, computational). The central noise model and the stochastic resonance model can be unified (b1). The homeostatic plasticity model is a specification of the central gain model (b2).} 
	\label{descriptionLevels_1}
\end{figure}

According to David Marr, one can seek to understand a system at complementary levels of analysis \cite{marr1979computational}. He distinguished the computational, the algorithmic and the implementational level of analysis (see Fig. \ref{descriptionLevels_1} a).
The \emph{computational level} is the most coarse-grained level of analysis. It asks what is the problem or task the system is seeking to solve via computation, that results in the observed phenomena; in our context, phantom perceptions like tinnitus. This level of analysis is addressed by the fields of psychology and cognitive neuroscience. In contrast, the \emph{implementational level} represents the most fine-grained description of a system. Here, the system’s concrete, physical layout is analyzed. In computer science and engineering, this corresponds to the exact hardware architecture and the individual software realization, with a particular programming language. In the brain, where there exists no clear distinction between software and hardware (or wetware), this level of description corresponds to the structural design of ion channels, synapses, neurons, local circuits and larger systems, and the physiological processes these components are subject to. This level of analysis can be considered as the hallmark of physiology and neurobiology. Finally, the \emph{algorithmic level} takes an intermediate position between the previously described levels. It is about which algorithms—that are physically realized at the implementational level—the system employs to manipulate its internal representations, in order to solve the tasks and problems identified at the computational level. In computer science, the algorithmic level would be described independently of a specific programming language by abstract pseudo code.

We argue that analysis at the algorithmic level is most crucial to understand auditory phantom perceptions like tinnitus or Zwicker tone. Only by knowing the algorithms that underlie normal auditory perception, we will gain a detailed understanding of what exactly happens under certain pathological conditions such as hearing loss. And which processes eventually cause the development of tinnitus, so that we can mitigate or reverse these processes. 

Which discipline addresses this level of analysis in tinnitus research? Computational neuroscience comes to mind immediately. However, in “good old-fashioned” computational neuroscience, great efforts have been made to model the physiological processes at the level of single neurons, dendrites, axons, synapses, or even ion channels, leading to increasingly complex computational models. These models, mostly based on systems of coupled differential equations, can mimic experimental data in great detail. Perhaps the most popular among these models is the famous Hodgkin-Huxley model \cite{hodgkin1952quantitative}, which reproduces the temporal course of the membrane potential of a single neuron with impressive accuracy. These types of models are of great importance to deepen our understanding of fundamental physiological processes. However, in our opinion, they must also be considered  as belonging to the implementational level of analysis, since they merely describe the physical realization of the algorithms, rather than the algorithms themselves.

In the following section, we will discuss emerging research directions that speak to the algorithmic level of analysis in the context of tinnitus research.

\subsection*{The integration of artificial intelligence in tinnitus research}

As we argued above, hypothesis testing alone does not lead to a mechanistic understanding. Instead, it needs to be complemented by the construction of task-pointing computational models, since only synthesis in a computer simulation can reveal the interaction of proposed components entailed by a mechanistic explanation, i.e., which algorithms are realized, and whether they can account for the perceptual, cognitive or behavioral function in question. As Nobel laureate and theoretical physicist Richard Feynman pointed out: \emph{``What I cannot create, I do not understand.''}

Along these lines, one may consider extending the four goals of psychology, i.e., to describe, explain, predict, and change cognition and behavior \cite{holt2019ebook}, by adding a fifth one: to build synthetic cognition and behavior.

As pointed out in previous publications \cite{marblestone2016toward, van2017computational, van2017artificial, barak2017recurrent, kriegeskorte2018cognitive}, these computational models can be based on constructs from artificial intelligence, for example deep learning \cite{lecun2015deep, schmidhuber2015deep}. A related development in artificial intelligence rests upon the explicit use of generative models, leading to formulations of action and perception, in terms of predictive coding and active inference. Examples of their application to auditory processing and hallucinations range from examining the role of certain oscillatory frequencies in message passing, through to simulations of active listening and speech perception. Please see \cite{benrimoh2018active,isomura2019bayesian,koelsch2019predictive,friston2021active,arnal2012cortical,hovsepyan2018combining,powers2017pavlovian} for details. Both deep learning and predictive processing accounts rest upon message passing on neural networks. 

Artificial deep neural networks are designed to solve problems clearly defined at the computational level of analysis, in our case auditory perception tasks like, e.g., speech recognition. These models are precisely defined at an algorithmic level, which is completely independent from any individual programming language or specific software library, i.e., the implementational level of analysis. Hence, these algorithms could, at least in principle, also be realized in the brain as biological neural networks. Once we have built such models and algorithms in computer simulations, we can subsequently compare their dynamics and internal representations with brain—and behavioral data—in order to reject or adjust putative models, thereby successively increasing biological fidelity \cite{kriegeskorte2018cognitive}. Vice versa, the ensuing models may also serve to generate new testable hypotheses about cognitive and neural processing in auditory neuroscience.

This research approach -- of combining artificial intelligence, cognitive science, and neuroscience -- has been coined as \emph{Cognitive Computational Neuroscience} (CCN) \cite{kriegeskorte2018cognitive}. Furthermore, besides the advantages discussed above, this approach furnishes the opportunity for in silico testing of new, putative treatment interventions for conditions like tinnitus, prior to in vivo experiments. In this way, CCN may even serve to reduce the number of animal experiments.

However, we note that Cognitive Computational Neuroscience of auditory perception is not only beneficial for neuroscience. As noted in \cite{hassabis2017neuroscience}, understanding biological brains could play a vital role in building intelligent machines, and that current advances in artificial intelligence have been inspired by the study of neural computation in humans and animals. Thus, Cognitive Computational Neuroscience of auditory perception, may contribute to the development of neuroscience-inspired artificial intelligence systems in the domain of natural language processing \cite{cambria2014jumping}. Finally, neuroscience may even serve to investigate \emph{machine behavior} \cite{rahwan2019machine}, i.e., illuminate the black box of deep learning  \cite{voosen2017ai, hutson2018artificial}. However, so far, most AI research these days does not even attempt to mimic or understand brain or biology in general.

In other neuroscientific strands, such as research on spatial navigation, the fusion of classical neuroscience and AI has already led to major breakthroughs, and still promises further advances in the future \cite{bermudez2020neuroscience}. For example, on the one hand Stachenfeld and colleagues developed a mathematical framework for the function of place and grid cells in the entorhinal-hippocampal system based on predictive coding \cite{stachenfeld2017hippocampus, mcnamee2021flexible}, on the other hand, researchers from Google DeepMind developed artificial agents based on Long-short-Term-Memory (LSTM, \cite{hochreiter1997long, hochreiter1997lstm}) neurons, in which place and grid cells emerged automatically \cite{banino2018vector}. In another AI model, Gerum and coworkers showed that spatial navigation in a maze could be achieved by very small neural networks, which are trained with an evolutionary algorithm and are evolutionary pruned \cite{gerum2020sparsity}.   

\section*{Towards a unified model of tinnitus development}

\subsection*{The hierarchy of the different tinnitus models}

In the following section, we describe a path towards a CCN of tinnitus research. Thus, in a first step we have to go back to Labzebnik \cite{lazebnik2002can} and find a way to communicate efficiently and formally about various tinnitus models. Extant tinnitus models can be sorted by the different levels of analysis according to Marr \cite{marr1979computational}, which means that each model can be assigned to one or more of the three categories (see Fig. \ref{descriptionLevels_1} b): implementational level (molecular mechanisms, synapses etc.), algorithmic level (how neural signals are translated to information processing), and computational level (what are the basic mathematical imperatives for processing, see also paragraph: What does it mean to understand a system?).

The three levels of analysis can be easily illustrated with the \emph{Lateral Inhibition Model} of tinnitus, which describes tinnitus as a result of decreased lateral inhibition (e.g. \cite{eggermont2003central, kral1996lateral}) due to decreased cochlear input; e.g., caused by a noised-induced cochlear synaptopathy \cite{tziridis2021tinnitus}. Thus, the lateral inhibition model explains tinnitus on all different levels of description. The implementational level (see Marrs level of analysis Fig. \ref{descriptionLevels_1} a), which corresponds to the molecular mechanisms of lateral inhibition, is nearly fully understood. For example, in the dorsal cochlear nucleus cartwheel cells release glycine to inhibit fusiform cells, which are excitatory \cite{roberts2010molecular, golding1997physiological, caspary2006age}. The computational role of inhibition is to narrow the input range of the fusiform cells \cite{roberts2010molecular}. To provide contrast enhancement via lateral inhibition, neurons surrounding a certain excitatory neuron, which receives auditory input, are inhibited. This wiring scheme ‘sharpens’ the tuning curves of neurons along the auditory pathway. The wiring scheme corresponds to the algorithmic level of analysis. The computational level of description is the mathematical description of decreased lateral inhibition. Thus, hearing loss leads to decreased input from the cochlea, which causes a decreased firing rate of the inhibitory neurons and thus to disinhibition of subsequent excitatory neurons. These properties can be easily written down in simple mathematical formulas. This means that the underlying mechanisms of the lateral inhibition model of tinnitus are fully understood from specific neurotransmitter processes to an abstract mathematical formulation. This is the goal of cognitive computational neuroscience. However, the fact that the model explains tinnitus manifestation on all scales does not say anything about the correctness of the model’s predictions. Indeed, a good model should be understood on all scales (implementational to computational), but it must also fit experimental observations, which is not the case in the \emph{Lateral Inhibition Model}. Other models trying to explain tinnitus do not provide the full explanatory power.

The thalamic bursting theory -- which proposes that bursting neurons in the thalamus cause tinnitus -- has a valid explanation for the origin of the spike bursts (low threshold calcium spikes, for details see \cite{jeanmonod1996low}). However, it remains elusive, in terms of how these bursts cause tinnitus. Other top-down models -- such as the predictive coding model \cite{sedley2016integrative}, based on the Bayesian Brain theory -- provide a valid mathematical description of the proposed mechanisms, but largely lack an explanation of how the Bayesian statistics can be implemented in a neural network and thus in the brain \cite{friston2012history}. However, there exist some first approaches toward neural networks for Bayesian inference which will ultimately prove possible, but are still not fully developed \cite{hawkins2004intelligence}. Other tinnitus models describe macro-phenomena such as the thalamo-cortical dysrhythmia \cite{llinas1999thalamocortical}, or describe tinnitus as a result of overlapping neural circuits \cite{de2011phantom}. Those models are phenomenological, but do not provide a mathematical description and thus are difficult to falsify or test in silico.


\subsection*{A critical role of stochastic resonance}

In the following paragraph we provide an in-depth discussion of central noise and central gain, as possible causes for tinnitus, and consider how to adjudicate between—or combine—these two theories. To discuss these two models and their relationship, it is necessary to introduce a proper nomenclature. Thus, in the following we refer to the mathematical description of Zeng, who describes central gain as a linear amplification factor $g$, which increases the input signal $I$. Central noise $N$ is a further additive term (see eq. \ref{eq_central_noise}) \cite{zeng2013active, zeng2020tinnitus}.
\begin{align}
s = \mathrm{\textbf {\textit g}} \cdot I + \mathrm{\textbf{\textit N}} 
\label{eq_central_noise}
\end{align}
Thus, also in most other studies central gain increase refers to an amplification of the input signal and is always associated with stimulus-evoked activity (for an extensive review see \cite{auerbach2014central}). In contrast to that, central noise is related to an increase of spontaneous activity in the absence of any external stimulation. Additionally, it is necessary to distinguish between the observable effect, i.e., increased central gain and central noise, and the underlying principles and mechanisms, e.g., homeostatic plasticity \cite{turrigiano1999homeostatic}, feedback loops \cite{brodal2004central} or stochastic resonance (SR) \cite{benzi1981mechanism,gammaitoni1998stochastic,mcdonnell2009stochastic}. In reality, the increased gain and noise cannot be fully decoupled, for example, an increased excitability of neurons along the auditory pathway caused by homeostatic plasticity automatically leads to an amplification of neural noise. Thus, the formula of the neural signal could be altered so that the amplification factor also has an effect on the central noise (mixing term: $g\cdot n$ in equation \ref{eq_mixing_term}). 
\begin{align}
s = g\cdot(I+N)
\label{eq_mixing_term}
\end{align}
Several neural mechanisms are considered here. First, homeostatic plasticity has been implicated for tinnitus generation (e.g. \cite{yang2011homeostatic}), however this mechanism is simply too slow to explain acute tinnitus phenomena after a noise trauma caused by a sudden loud stimulus \cite{axelsson1987acute}. 

In contrast, neural circuits operating on faster time scales can explain acute tinnitus: namely, tinnitus is caused by a sub-cortical feedback loop adapting neural noise input into the auditory system \cite{krauss2017adaptive}. Here we suggest that stochastic resonance plays a critical role in not only generating tinnitus but also restoring hearing \cite{krauss2016stochastic,krauss2017adaptive,schilling2021stochastic}. To illustrate this role, we rewrite equation \ref{eq_mixing_term} in a classical signal detection task, in which the neural signal has to reach a threshold So for the input signal $I$ to be detected:
\begin{align}
s = g\uparrow\cdot(I\downarrow+N\uparrow)
\label{eq_mixing_term_HL}
\end{align}
In cases of hearing loss, the input $I$ is effectively reduced. Therefore, to reach the same neural threshold, one could either increase the central noise $N$, or the central gain $G$, or both. Because increasing gain results in a squared increase in variance \cite{zeng2020tinnitus}, which increases the difficulty of signal detection, it is not the most economical means of compensating for hearing loss in cognitive neural computation (e.g., Occam’s razor). Instead, it makes sense to add internal neural noise to lift weak input signals above the sensory threshold, a mechanism known as stochastic resonance (SR) \cite{benzi1981mechanism,gammaitoni1998stochastic,mcdonnell2009stochastic}. In traditional SR, a nonlinear device such as hard thresholding, and periodic signals are needed \cite{gammaitoni1998stochastic}. Recently, it has been shown that auto-correlation can serve as an estimator for the information content of the signal, even if it is non-periodic \cite{krauss2017adaptive}.

The critical role of stochastic resonance is supported by broad empirical evidence:

First, additional intrinsic neural noise \cite{krauss2016stochastic, gollnast2017analysis} as well as external acoustic noise \cite{zeng2000human} can improve pure-tone hearing thresholds by approximately 5\,dB. However, this 5\,dB threshold decrease (i.e., improvement) does not explain why this mechanism is evolutionary advantageous, as the cost of a potentially annoying and morbid tinnitus perception may be high. In a computational model, it has been showed that frequency-specific intrinsic neural noise has the potential to significantly improve speech recognition by a far larger amount (up to a factor of 2) \cite{schilling2020intrinsic}. This improvement in speech comprehension and the perception of complex sounds -- which could be also important for orienting animals as warning sounds -- could be an explanation for the emergence of this mechanism in our auditory system during evolution. In recent studies, the fact that different modalities use SR to improve the signal has been proven \cite{platerremote, yashima2021auditory}. It seems that SR and especially cross-modal SR is a universal principle of sensory processing \cite{krauss2018cross}.


Second, central noise is needed to stabilize a biological system. Zeng showed that \emph{``mathematically, the loudness at threshold is infinite when the internal noise is zero (c = 0), and vice versa. This is a fundamental argument for why the brain has or needs internal noise because infinite loudness is clearly biologically unacceptable''} \cite{zeng2020unified}.

Third, the central noise model based on the SR mechanism provides a mechanistic explanation for the purpose of the somatosensory projections to auditory nuclei such as the DCN \cite{dehmel2012noise, shore2006somatosensory, wu2016tinnitus}. Actually, very recently, Koops and Eggermont argued that \emph{``increased and uncorrelated noise, potentially the result from a noise source outside of the auditory pathway''} \cite{koops2021thalamus} might play a major role in tinnitus development. Potentially, this somatosensory input is nothing else than intrinsically generated neural noise, which is modulated in the DCN to leverage SR in the auditory system. This theory accords with the finding that tinnitus can be modulated by somatosensory input like, e.g., jaw movement \cite{pinchoff1998modulation, lanting2010neural, won2013prevalence}. Furthermore, tinnitus development can be prevented \cite{sturm2017noise} or suppressed \cite{schaette2010acoustic, schilling2020reduktion,tziridis2022spectrally} by the presentation of external acoustic noise, which works best when the noise spectrum covers the impaired frequencies and the tinnitus pitch \cite{schaette2010acoustic, schilling2020reduktion,tziridis2022spectrally}. In a recent study, a novel approach has been developed combining somatosensory stimulation with auditory stimulation, in order to modulate the tinnitus loudness \cite{conlon2020bimodal}. Finally, it has been demonstrated that electro-tactile stimulation of the finger tips enhances cochlear implant speech recognition in noise \cite{huang2017electrospeech}, Mandarin tone recognition \cite{huang2017electroMandarin}, and melody recognition \cite{huang2020electro}. While the authors did not make any mention of stochastic resonance or internal noise, it is a reasonable assertion that the observed effect might have acted via cross-modal stochastic resonance \cite{krauss2018cross}.

\begin{figure}[ht!]
	\centering
	\includegraphics[width = \textwidth]{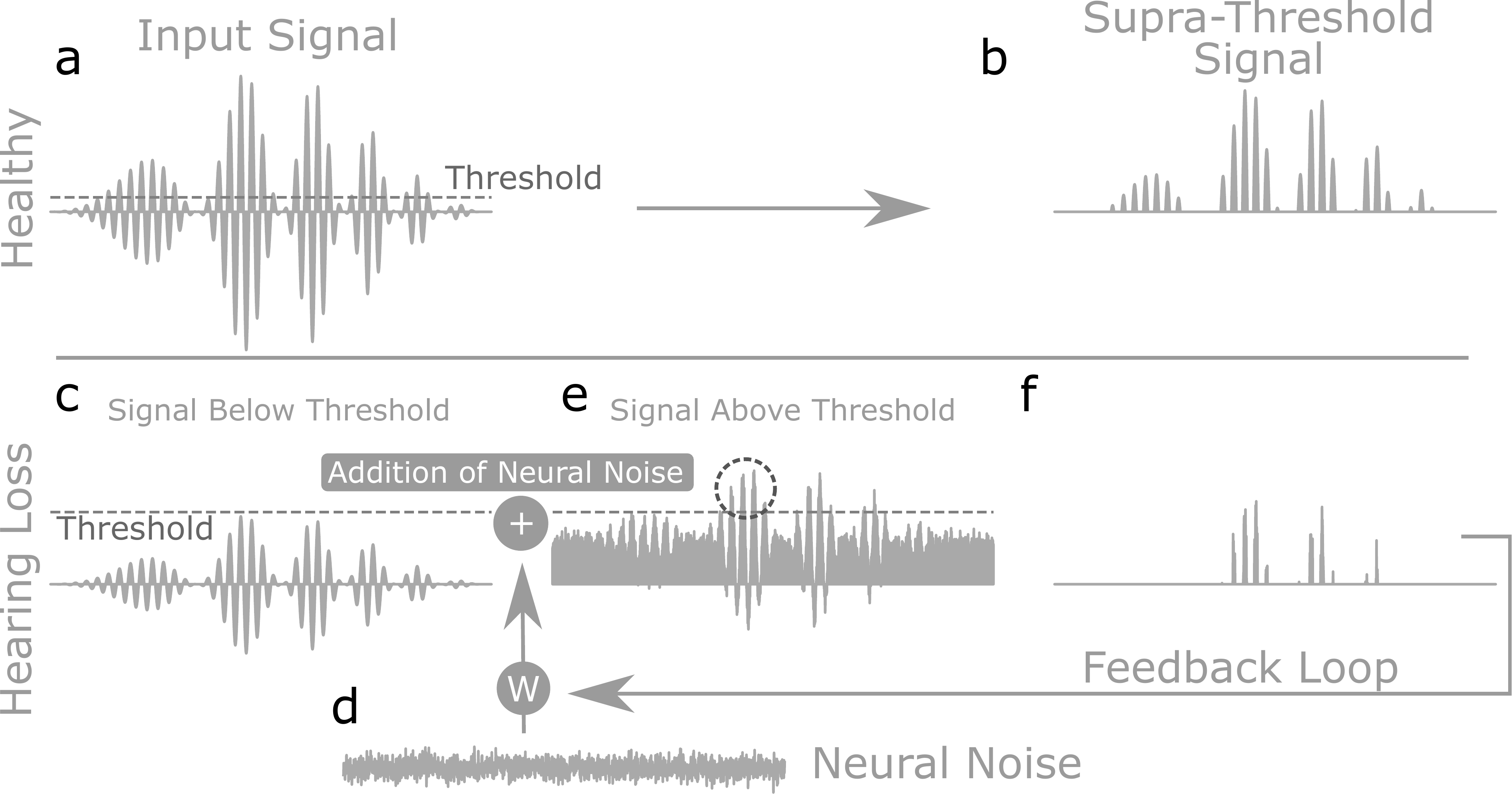}
	\caption{\textbf{Stochastic resonance model of tinnitus induction:} In the healthy auditory system, the input signal (a) can pass the detection threshold resulting in a supra-threshold signal as output (b). In case of hearing loss, the input signal remains below the threshold (c), resulting in zero output. However, if the optimum amount neural noise (d) is added to the weak signal, then signal plus noise can pass the threshold again (e), making a previously undetectable signal, detectable again (f). The optimum amount of noise depends on the momentary statistics of the input signal, and is continuously adjusted via a feed-back loop. This processing principle is called adaptive stochastic resonance.} 
	\label{SR_Tinnitus}
\end{figure}

\subsection*{Tinnitus and the Bayesian brain}

\begin{figure}[ht!]
	\centering
	\includegraphics[width = \textwidth]{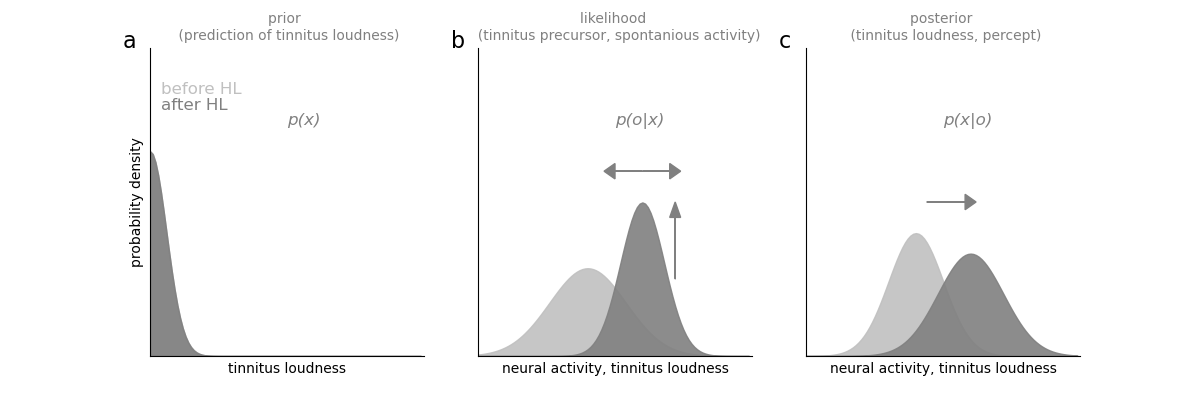}
	\caption{\textbf{Predictive coding model of tinnitus induction:} a:	Prior: the probability of a certain percept ($x$: loudness); b: Likelihood: The probability that an auditory observations o leads to a certain auditory percept $x$; c: The posterior corresponds to the perceived loudness. A hearing loss (dark gray: after hearing loss, light gray: healthy condition) leads to an increased precision and mean of the likelihood, which leads to an increased mean of the posterior and thus to a tinnitus percept.} 
	\label{Bayesian_Brain}
\end{figure}

The central gain and the central noise model provide a sophisticated and mathematically well-developed explanation for the tinnitus-related neural hyperactivity in the brainstem. However, these theories do not explain why this hyperactivity is transmitted through the thalamus and induces a conscious experience. Furthermore, the models do not make predictions on tinnitus heterogeneity. In particular, tinnitus is probably always caused by hearing loss, but hearing loss does not necessarily lead to tinnitus \cite{tan2013tinnitus}. The only model with a solid mathematical background dealing with this issue is the sensory precision model from Sedley et al. \cite{sedley2016integrative} which is based on the algorithmic formulation of predictive coding within the computational “Bayesian brain” hypothesis \cite{knill2004bayesian, friston2010free, friston2012history, bastos2012canonical, clark2013whatever, de2014bayesian}. Bayesian formulations of predictive processing are based on the Bayes theorem (eq. \ref{eq_bayes}, \cite{vilares2011bayesian, stigler2013true}) that describes, mathematically, how to update beliefs in the light of new incoming information.
\begin{align}
    p(x|o) \sim p(o|x) \, p(x)
    \label{eq_bayes}
\end{align}
In our context, the brain is continuously updating its posterior belief distribution $p(x|o)$ ) about the actual sound intensity $x$, given auditory afferents or observations $o$. This posterior belief is presumably correlated with the auditory perception and subjective loudness. This update is achieved by combining the prior expectations $p(x)$, descending from the higher regions of the processing hierarchy, with precision weighted prediction errors ascending from below, which mathematically report the likelihood $p(o|x)$ (cf. Fig. \ref{Bayesian_Brain}). ‘Likelihood’ refers to the probability that the pattern of sensory input indicates a particular underlying sensory event (such as a specific word being spoken). In predictive coding, Bayesian belief updating—from prior to posterior beliefs (cf. Equation \ref{eq_bayes}) -- are driven by prediction errors; namely, sensory information that is newsworthy, in the sense that it has not been predicted. The precision weighting reflects the precision of the likelihood; that is, a precise likelihood mapping means that sensory prediction errors are more informative and therefore produce more belief updating higher in the auditory hierarchy. Physiologically, this is usually associated with an increase in the postsynaptic gain or excitability of neuronal populations reporting prediction errors (usually superficial pyramidal cells in the cortex): see \cite{benrimoh2018active,friston2017graphical,kanai2015cerebral,shipp2016neural,adams2013computational,sterzer2018predictive,bastos2012canonical} for a predictive coding account of neuronal message passing and their role in hallucinatory phenomena.

According to Sedley and coworkers \cite{sedley2016integrative}, even in a silent environment, lower levels of the auditory hierarchy report only low amplitude auditory signals, which are relatively compatible with the default hypothesis of ‘silence’ (i.e., large posterior probabilities for low sound intensities). In normal hearing, these ‘tinnitus precursors’ are ignored by attenuating their precision -- a process sometimes associated with sensory attenuation. However, if the brain recognizes sensory de-afferentation due to hearing loss, it may assign more precision to ascending auditory prediction errors (e.g. through increased central gain), thereby falsely interpreting tinnitus precursors as an auditory percept. In short, a failure of sensory attenuation means that ‘silence’ is misinterpreted as ‘sound’ -- and tinnitus ensues \cite{sedley2016integrative}. 

However, this Bayesian account makes no specific claims about the neural mechanisms that mediate changes in precision or the tinnitus precursor. To furnish a complete mechanistic explanation, a process theory, such as the central noise model, is needed. For example, the predictive coding model suggests that aberrant precision (i.e., synaptic gain control) leads to a misinterpretation of auditory afferents such that ‘silence’ is misinterpreted as a phantom ‘sound’. This computational account can now be supplemented with the process theory afforded by central noise gain. In brief, central noise augments postsynaptic gain—via stochastic resonance—to adaptively respond to the loss of precise sensory afferents. At the same time, the requisite central noise augments -- or supplants -- the ‘tinnitus precursor’. On this view, the central noise model equips a computational (Bayesian inference) account with an implementational (stochastic neuronal resonance) mechanism that is consistent with an algorithmic (predictive coding) formulation of auditory processing and tinnitus.

\section*{Conclusion and Outlook}

\begin{figure}[ht!]
	\centering
	\includegraphics[width=8.5cm]{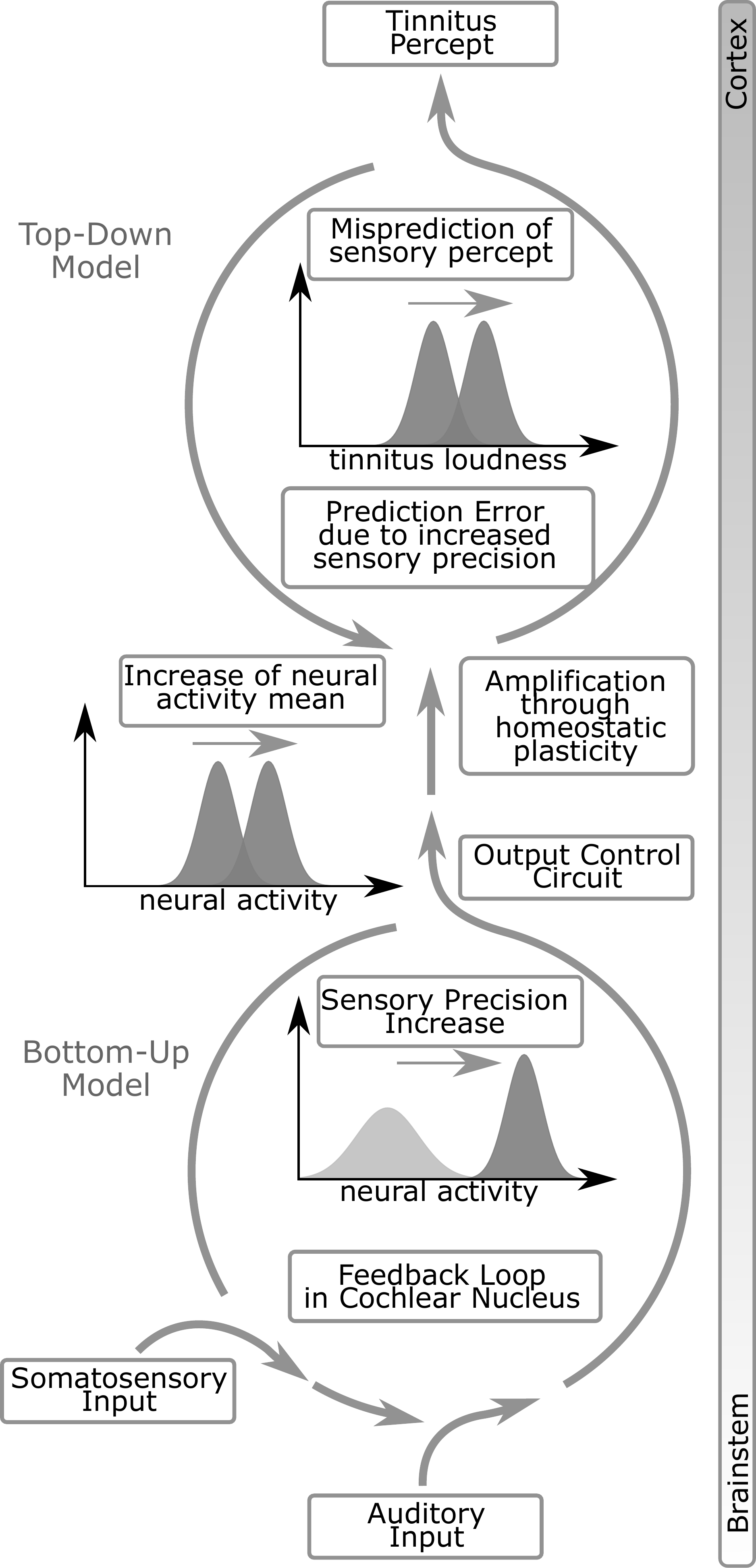}
	\caption{\textbf{Unified theory of tinnitus development:} The unified model can be divided into two main parts: the bottom-up model (central noise and SR) and the top-down model (predictive coding). Implementation Model: Decreased input from the cochlea due to a hearing loss is compensated by the addition of intrinsically generated neural noise (central noise) from the somatosensory system, which is used to restore hearing through the exploitation of the stochastic resonance phenomenon and accompanying increase in postsynaptic gain or sensitivity in the auditory hierarchy. The optimal neural noise level is constantly tuned via a feedback loop, which maximizes information transmission, quantifiable as auto-correlation of the input signal. This feedback loop increases the average noise amplitude and decreases the variance of the noise distribution (increased sensory precision). The noisy signal is further propagated upwards and, after cochlear damage, additionally amplified along the auditory pathway by homeostatic plasticity. The amplified noisy signal is the input to a second feedback loop in higher brain areas such as the auditory cortex. Algorithmic model: The descending part of this second loop predicts the auditory input (prior), whereas the ascending part contains the prediction error (likelihood). The persistent mis-prediction of the auditory input prevents the thalamus from suppressing the signal, hence it becomes conscious and is finally perceived as tinnitus.} 
	\label{merged_model}
\end{figure}

In conclusion, the combination of the process theory of central noise increase and adaptive stochastic resonance -- as a bottom-up mechanism -- together with the computational model of predictive coding -- as a complementary top-down mechanism -- provides an integrated explanation of tinnitus induction as well as its heterogeneity (cf. Figure \ref{merged_model}). Furthermore, the models provide a mathematical framework, which can be used to make quantitative predictions that can be tested through novel experimental paradigms. Since both SR and predictive coding as universal processing mechanisms are ubiquitous in the brain, we speculate that the presented integrative framework may extend to the perception of other sensory modalities and even beyond to certain aspects of cognition and behavior in general.

A current challenge is a network theory of predictive coding, which explains how these computations are implemented in the brain \cite{friston2012history}. See \cite{friston2017graphical,shipp2016neural,bastos2012canonical,adams2013predictions,da2021neural,friston2017active} for attempts to place predictive coding in the larger context of Bayesian belief updating in the brain.

Our integrated model of auditory (phantom) perception demonstrates that the fusion of computational neuroscience, AI, and experimental neuroscience leads to innovative ideas and paves the way for further advances in neuroscience and AI research. For instance, novel evaluation techniques for neurophysiological data based on AI and Bayesian statistics were recently established \cite{krauss2018statistical, krauss2018analysis, krauss2021analysis, metzner2021sleep}, the role of noise  in neural networks and other biological information processing systems was considered in  \cite{krauss2017chemical, krauss2019recurrence, metzner2021dynamical, bonsel2021control}, and the benefit and application of noise and randomness in Machine Learning approaches was further investigated in \cite{schilling2020intrinsic, yang2021neural, harikrishnan2021noise}. 
On the one hand, the fusion of these complementary fields may evince the neural mechanisms of tinnitus (CCN, \cite{kriegeskorte2018cognitive}) and information processing principles that underwrite functional brain architectures. On the other hand (neuroscience-inspired AI, \cite{hassabis2017neuroscience}) may accelerate research in machine learning. We hope that the four major steps towards a CCN of tinnitus, i.e. (i) \emph{finding an exact language}, (ii) \emph{developing a mechanistic theory}, (iii) \emph{testing the methods in fully specified test systems}, and (iv) \emph{merging AI with computational and experimental neuroscience}, will afford novel opportunities in tinnitus research. 

\FloatBarrier
\newpage
\section*{Acknowledgments}

This work was funded by the Deutsche Forschungsgemeinschaft (DFG, German Research Foundation): grant KR5148/2-1 (project number 436456810) to PK, and grant SCHI\,1482/3-1 (project number 451810794) to AS. Additionally, PK was supported by the Emerging Talents Initiative (ETI) of the University Erlangen-Nuremberg (grant 2019/2-Phil-01), and KF is supported by funding for the Wellcome Centre for Human Neuroimaging (Ref: 205103/Z/16/Z) and a Canada-UK Artificial Intelligence Initiative (Ref: ES/T01279X/1). Furthermore, the research leading to these results has received funding from the European Research Council (ERC) under the European Union’s Horizon 2020 research and innovation programme (ERC Grant No. 810316 to AM). Finally, we wish to thank Arnaud Norena for useful discussion.

\section*{Author contributions}
All authors discussed the cited studies and their interpretation, developed the presented unified model, and wrote the manuscript.

\section*{Competing interests}
The authors declare no competing financial interests.

\FloatBarrier

\end{document}